\documentclass[10 pt, letterpaper]{article}

\usepackage[T1]{fontenc}
\usepackage[latin9]{inputenc}
\usepackage{color}
\usepackage{amsmath}
\usepackage{amssymb}
\usepackage{graphicx}

\makeatletter

\usepackage{makeidx}\usepackage{times}

\newcommand{\Z}{\mathbb{Z}}
\newcommand{\R}{\mathbb{R}}
\newcommand{\N}{\mathbb{N}}
\newtheorem{definition}{Definition}
\newtheorem{assumption}{Assumption}
\newtheorem{theorem}{Theorem}
\newtheorem{corollary}{Corollary}

\newtheorem{procedure}{Procedure}
\newtheorem{algorithm1}{Algorithm}

\begin{document}
\date{}

\title{Identification of nonlinear controllers from data:\\ theory and computation\thanks{This research has received funding from the California Energy Commission under the EISG grant n. 56983A/10-15 ``Autonomous flexible wings for high-altitude wind energy generation'', and from the European Union Seventh Framework Programme (FP7/2007-2013) under grant agreement n. PIOF-GA-2009-252284 - Marie Curie project ``Innovative Control, Identification and Estimation Methodologies for Sustainable Energy Technologies''.}}
\author{L. Fagiano\thanks{L. Fagiano is with the Automatic Control Laboratory, Swiss Federal Institute of Technology,Zurich, Switzerland. E-mail: fagiano@control.ee.ethz.ch.}$\,$ and  C. Novara\thanks{C. Novara is Dipartimento di Automatica e Informatica, Politecnico di Torino,  Torino - Italy. E-mail: carlo.novara@polito.it.}}
\maketitle

\section{Introduction}
This manuscript contains technical details and proofs of recent results developed by the authors, pertaining to the design of nonlinear controllers from the experimental data measured on an existing feedback control system.

\section{Problem formulation}

\label{S:problem} 
The setting we consider in this work is the following. A single-input,
discrete time, nonlinear dynamical system of interest operates in
closed loop with an existing controller. Both the system and the controller
are not known. The system's input variable $u(t)$, i.e. the controller's
output, is known and it can be measured at discrete time instants
$t\in\Z$. Moreover, $u$ is limited in a compact $U=[\underline{u},\overline{u}]$.
The system's output variable $y(t)$, i.e. the controller's input,
is not known a priori but the control designer can rely on sensors
to acquire measurements of different ``candidate'' feedback variables,
based on her/his intuition and experience with the physical process
under study. The output $y$ is assumed to belong to a compact set
$Y\subset\R^{n_{y}}$. After a choice of $y(t)$ has been made, we
assume that the controller is a static function of this variable:
\begin{equation}
\begin{array}{l}
u(t)=\kappa(y(t))\\
\kappa:Y\rightarrow U.
\end{array}\label{E:controller}
\end{equation}
Moreover, we assume that a disturbance variable $e_{s}(t)$ is acting
on the dynamical system. The variable $e_{s}$ accounts for (a) exogenous
disturbances, (b) neglected and time-varying dynamics, and (c) the
approximation error induced by choosing the input of the controller
to be equal to $y$. The value of $e_{s}(t)$ is also assumed to belong
to a compact set $E_{s}\subset\R^{n_{e}}$. We then assume that the
chosen output variable evolves in time as follows:
\begin{equation}
\begin{array}{l}
y(t+1)=f(y(t),u(t),e_{s}(t))\\
f:Y\times U\times E_{s}\rightarrow Y.
\end{array}\label{E:sys2}
\end{equation}

Let us now introduce three assumptions on functions $f$ and $\kappa$.
In the following, we will make use of the function sets $\mathcal{K}$
and $\mathcal{KL}$: to this end, we recall that $\mathcal{K}$ is
the set of all strictly increasing functions $\alpha:\R^{+}\rightarrow\R^{+}$
such that $\alpha(0)=0$, while $\mathcal{KL}$ is the set of all
functions $\beta:\R^{+}\times\R^{+}\rightarrow\R^{+}$ such that for
fixed $t$, $\beta(x,t)\in\mathcal{K}$, and for fixed $x$, $\lim\limits _{t\rightarrow\infty}\beta(x,t)=0$.

\begin{assumption}\label{A:system_cont} The function $f$ is Lipschitz
continuous over the compact $Y\times U\times E_{s}$. In particular,
it holds that
\begin{equation}
\begin{array}{l}
\exists\gamma_{f}\in(0,+\infty):\forall e_{s}\in E_{s},\forall y\in Y,\,\forall u^{1},\, u^{2}\in U,\,\|f(y,u^{1},e_{s})-f(y,u^{2},e_{s})\|_{\infty}\leq\gamma_{f}|u^{1}-u^{2}|.\end{array}\label{E:unif_cont_f}
\end{equation}

\hfill{}$\square$

\end{assumption}

\begin{assumption}\label{A:controller_cont} The function $\kappa$
is Lipschitz continuous over the compact $Y$.\hfill{}$\square$

\end{assumption}

Assumptions \ref{A:system_cont}-\ref{A:controller_cont} imply that
the closed loop system:
\begin{equation}
\begin{array}{l}
y(t+1)=g(y(t),e_{s}(t))\doteq f(y(t),\kappa(y(t)),e_{s}(t))\\
g:Y\times E_{s}\rightarrow Y
\end{array}\label{E:closed_loop}
\end{equation}
is also described by a Lipschitz continuous function $g$. In particular,
by construction, the function $g$ enjoys the following properties:
\begin{equation}
\begin{array}{l}
\exists\gamma_{g,y}\in(0,+\infty):\forall e_{s}\in E_{s},\forall y^{1},\, y^{2}\in Y,\,\|g(y^{1},e_{s})-g(y^{2},e_{s})\|_{\infty}\leq\gamma_{g,y}\|y^{1}-y^{2}\|_{\infty},\end{array}\label{E:unif_cont_y}
\end{equation}
\begin{equation}
\begin{array}{l}
\exists\gamma_{g,e}\in(0,+\infty):\forall y\in Y,\forall e_{s}^{1},\, e_{s}^{2}\in E_{s},\,\|g(y,e_{s}^{1})-F(g,e_{s}^{2})\|_{\infty}\leq\gamma_{g,e}\|e_{s}^{1}-e_{s}^{2}\|_{\infty}.\end{array}\label{E:unif_cont_e}
\end{equation}
Assumptions \ref{A:system_cont}-\ref{A:controller_cont} are quite
standard in nonlinear control analysis and design and they are reasonable,
since in practice the inputs, disturbance and outputs of the process
under study are often bounded in some compact sets and the functions
describing the system and the controller are assumed to be differentiable
on such compact sets, hence Lipschitz continuous.

The dynamical system described by $g$ has $e_{s}$ as input and $y$
as output. We denote with $g_{0}\doteq g(0,0)$ the value of $g$
evaluated at $y=0$, $e_{s}=0$. The properties of the closed-loop
system clearly depend on the controller $\kappa$, which is assumed
to be stabilizing. In particular, we consider the following notion
of stability:

\begin{definition}\label{D:stability} A nonlinear system with input
$e_{s}$ and output $y$, is finite-gain $\ell_{\infty}$ stable if
a function $\alpha\in\mathcal{K}$, a function $\beta\in\mathcal{KL}$
and a scalar $\delta>0$ exist, such that:
\begin{equation}
\forall t\geq0,\,\|y(t)\|_{\infty}\leq\alpha(\|\textbf{e}_{s}\|_{\infty})+\beta(\|y(0)\|_{\infty},t)+\delta.\label{E:stability_def}
\end{equation}

\hfill{}$\square$

\end{definition}

In Definition \ref{D:stability}, the generic signal $\textbf{v}\doteq\{v(0),v(1),...\}$
is given by the infinite sequence of values of the variable $v(t),\, t\geq0,$
and $\|\textbf{v}\|_{\infty}\doteq\max\limits _{t\geq0}\|v(t)\|_{\infty}$
is the $\ell_{\infty}-$norm of the signal $\textbf{v}$ with the
underlying norm taken to be the vector $\infty-$norm $\|v\|_{\infty}$.\\
 The stabilizing properties of $\kappa$ are formalized by the following
assumption:

\begin{assumption}\label{A:controller_stab} The functions $\kappa$
and $f$ are such that property (\ref{E:unif_cont_y}) holds with
$\gamma_{g,y}(x)<1$.\hfill{}$\square$

\end{assumption}

Assumption \ref{A:controller_stab} implies that the closed-loop system
(\ref{E:closed_loop}) enjoys finite-gain $\ell_{\infty}$ stability
as given in Definition \ref{D:stability}, in particular we have:
\begin{equation}
\forall t\geq0,\,\|y(t)\|_{\infty}\leq\underbrace{\dfrac{\gamma_{g,e}}{1-\gamma_{g,y}}\|\textbf{e}_{s}\|_{\infty}}\limits _{\alpha(\|\textbf{e}_{s}\|_{\infty})}+\underbrace{\gamma_{g,y}^{t}\|y(0)\|_{\infty}}\limits _{\beta(\|y(0)\|_{\infty},t)}+\underbrace{\dfrac{1}{1-\gamma_{g,y}}\|g_{0}\|_{\infty}}\limits _{\delta},\label{E:stability_kappa}
\end{equation}
see the Appendix for a derivation of this inequality.

Overall, Assumptions \ref{A:system_cont}-\ref{A:controller_stab}
are quite common in the context of system identification, function
approximation and learning, since a stable system is needed to collect
data and carry out identification experiments. In particular, in this
work we will consider a finite number $N$ of input and output measurements,
indicated as $\tilde{u}(k),\,\tilde{y}(k),\, k=0,\ldots,N-1$, collected
from the system operating in closed loop with the unknown controller
$\kappa$. These data points are assumed to be affected by additive
noise variables, indicated as $e_{u}(t)$ and $e_{y}(t)$, respectively:
\begin{equation}
\begin{array}{l}
\tilde{u}(t)=u(t)+e_{u}(t)\\
\tilde{y}(t)=y(t)+e_{y}(t).
\end{array}\label{E:measures}
\end{equation}
Note that $e_{u}(t)$ may include both measurement noise and errors
arising in the application of the control law. The latter can be present
for example if the aim is to learn a controller from the behavior
of a human operator, who might be subject to fatigue and mistakes.

The noise variables are assumed to satisfy the following boundedness
properties where, for a generic variable $q\in\R^{n_{q}}$ and scalar
$\rho\in(0,+\infty)$, we denote the $n_{q}-$dimensional $\infty$-norm
ball set of radius $\rho$ as $B_{\rho}\doteq\{q\in\R^{n_{q}}:\|q\|_{\infty}\leq\rho\}$:

\begin{assumption}\label{A:noise} The following boundedness properties
hold:
\begin{description}
\item [{(a)}] $e_{u}(t)\in B_{\varepsilon_{u}},\,\forall t\geq0$ ;
\item [{(b)}] $e_{y}(t)\in B_{\varepsilon_{y}},\,\forall t\geq0$ .\hfill{}$\square$
\end{description}
\end{assumption}

According to (\ref{E:controller}), with straightforward manipulations,
the measured data can be described by the following set of equations:
\[
\tilde{u}(k)=\kappa(\widetilde{y}(k))+d\left(k\right),\, k=0,\ldots,N-1
\]
where $d\left(k\right)$ accounts for the noises $e_{u}(t)$ and $e_{y}(t)$
in (\ref{E:measures}). Since $e_{u}(t)$ and $e_{y}(t)$ are bounded
and $\kappa$ is Lipschitz continuous, it follows that $d\left(k\right)$
is also bounded:
\begin{equation}
d\left(k\right)\in B_{\varepsilon},\:\forall k\geq0.\label{eq:zi_bound}
\end{equation}
The following assumption on the pair $\left(\widetilde{y}(k),d\left(k\right)\right)$
is considered.

\begin{assumption} \label{dense_ass} The set of points $\mathcal{D}_{yd}^{N}\doteq\left\{ \left(\widetilde{y}(k),d\left(k\right)\right)\right\} _{k=0}^{N-1}$
is dense on $Y\times B_{\varepsilon}$ as $N\rightarrow\infty$. That
is, for any $(y,d)\in Y\times B_{\varepsilon}$ and any $\lambda\in\R^{+}$,
a value of $N_{\lambda}\in\N,\, N_{\lambda}<\infty$ and a pair $\left(\widetilde{y}(k),d\left(k\right)\right)\in\mathcal{D}_{yd}^{N_{\lambda}}$
exist such that $\left\Vert (y,d)-\left(\widetilde{y}(k),d\left(k\right)\right)\right\Vert _{\infty}\leq\lambda.$\hfill{}$\square$

\end{assumption} 
Assumption \ref{dense_ass} essentially ensures that the controller
domain $Y$ is ``well explored\textquotedblright{} by the data $\widetilde{y}(k)$
and, at the same time, the noise $d(k)$ covers its domain $B_{\varepsilon}$,
hitting the bounds $-\varepsilon$ and $\varepsilon$ with arbitrary
closeness after a sufficiently long time. This latter noise property
is called tightness, see \cite{NoFaMiAUT13} and, for a probabilistic
version, \cite{BCT98}.

In the described setting, the problem we want to address can be stated
as follows:\medskip{}

\textbf{Problem} 1: learn a controller $\hat{\kappa}$ from $N$ measurements
$\tilde{y}$ and $\tilde{u}$, obtained from the system operating
in closed-loop with an unknown controller $\kappa$, such that:
\begin{enumerate}
\item asymptotically, i.e. as $N\rightarrow\infty$, $\hat{\kappa}$ renders
the closed loop system finite-gain $\ell_{\infty}$ stable;
\item the trajectory deviation induced by the use of $\hat{\kappa}$ instead
of $\kappa$ is ``small'';
\item $\hat{\kappa}$ has ``low'' complexity, to be easily implementable
on real-time processors.\hfill{}$\square$ \medskip{}

\end{enumerate}

\section{Theoretical results and computation}

\label{S:stability}In this section, we present an approach that is
able to solve \textbf{Problem} 1. In order to do so, we first derive
a sufficient condition for a generic controller $\hat{\kappa}\approx\kappa$
to stabilize the closed-loop system and then we propose a technique,
based on convex optimization, that is able to learn a controller $\hat{\kappa}$
which enjoys asymptotically the derived stability condition.

\subsection{Closed loop stability analysis}

\label{SS:stability} Our first aim is to derive a sufficient condition
on the controller $\hat{\kappa}$, such that the obtained closed loop
system is finite-gain $\ell_{\infty}$ stable. The controller $\hat{\kappa}$
is chosen to be a Lipschitz continuous function over the compact $Y$,
with constant $\gamma_{\hat{\kappa}}$:
\begin{equation}
\exists\gamma_{\hat{\kappa}}\in(0,+\infty):\forall y^{1},y^{2}\in Y,\,|\hat{\kappa}(y^{1})-\hat{\kappa}(y^{2})|\leq\gamma_{\hat{\kappa}}\|y^{1}-y^{2}\|_{\infty}.\label{E:approx_lipschitz}
\end{equation}
Let us define the error function $\Delta:Y\rightarrow\R$:
\begin{equation}
\Delta(y)\doteq\kappa(y)-\hat{\kappa}(y).\label{E:error_function}
\end{equation}
We denote with $\Delta_{0}\doteq\Delta(0)$ the error function evaluated
at $y=0$. By construction, the error function is Lipschitz continuous,
with some constant $\gamma_{\Delta}\in(0,+\infty)$:
\begin{equation}
\exists\gamma_{\Delta}\in(0,+\infty):\forall y^{1},y^{2}\in Y,\,|\Delta(y^{1})-\Delta(y^{2})|\leq\gamma_{\Delta}\|y^{1}-y^{2}\|_{\infty}.\label{E:error_lipschitz}
\end{equation}
We indicate with $\hat{g}$ the closed loop system obtained by using
the controller $\hat{\kappa}$. In particular, $\hat{g}$ is defined
as follows:
\begin{equation}
\begin{array}{l}
y(t+1)=\hat{g}(y(t),e_{s}(t),e_{y}(t))\doteq f(y(t),\hat{\kappa}(y(t)+e_{y}(t)),e_{s}(t))\\
\hat{g}:Y\times E\times B_{\varepsilon_{y}}\rightarrow Y.
\end{array}\label{E:closed_loop_approx}
\end{equation}
Note that the feedback variable used by the learned controller $\hat{\kappa}$
is the noise-corrupted measurement of the output $y$. The next result
provides a sufficient condition for the controller $\hat{\kappa}$
to stabilize the closed loop system.

\begin{theorem}\label{T:stability} Let Assumptions \ref{A:system_cont}-\ref{A:controller_stab}
and \ref{A:noise}-(b) hold. If
\begin{equation}
\gamma_{\Delta}<\dfrac{1-\gamma_{g,y}}{\gamma_{f}},\label{E:stability_cond}
\end{equation}
then the closed-loop system $\hat{g}$ is finite-gain $\ell_{\infty}$
stable. More precisely, it holds that
\begin{equation}
\forall t\geq0,\,\|y(t)\|_{\infty}\leq\underbrace{\dfrac{\gamma_{g,e}}{1-\gamma}\|\textbf{e}_{s}\|_{\infty}}\limits _{\alpha(\|\textbf{e}_{s}\|_{\infty})}+\underbrace{\gamma^{t}\|y(0)\|_{\infty}}\limits _{\beta(\|y(0)\|_{\infty},t)}+\underbrace{\dfrac{1}{1-\gamma}\left(\|g_{0}\|_{\infty}+\gamma_{f}|\Delta_{0}|+\gamma_{f}\gamma_{\hat{\kappa}}\varepsilon_{y}\right)}\limits _{\delta},\label{E:stability_kappa_hat}
\end{equation}
with $\gamma\doteq(\gamma_{\Delta}\gamma_{f}+\gamma_{g,y})<1$. \end{theorem}
\emph{Proof.} See the Appendix.\hfill{}$\square$

It is worth commenting on the result of Theorem \ref{T:stability}.
Roughly speaking, the quantity $\gamma_{\Delta}$ gives an indication
on the regularity of the error function $\Delta=\kappa-\hat{\kappa}$.
Assuming for example that $\Delta$ is differentiable, a low $\gamma_{\Delta}$
means that the quantity $\|d\Delta/dy\|$ is bounded by a small value,
i.e. the variability of the error over the set $Y$ is low. This happens
e.g. when the functions $\kappa$ and $\hat{\kappa}$ differ by some
offset, but have similar shapes. A large value of $\gamma_{\Delta}$,
on the other hand, indicates that the control error can have high
variability, as it can happen e.g. when the controller $\hat{\kappa}$
is over-fitting the available measured data. Theorem \ref{T:stability}
states that the quantity $\gamma_{\Delta}$ should be sufficiently
small in order to guarantee closed-loop stability, and how small depends
on the features of the plant to be controlled and of the unknown controller
$\kappa$. In particular, the more ``sensitive'' is the plant to
input perturbation, i.e. the larger is the Lipschitz constant $\gamma_{f}$,
and the worse are the stabilizing properties of the controller $\kappa$,
i.e. the closer is the Lipschitz constant $\gamma_{g,y}$ is to 1,
the smaller $\gamma_{\Delta}$ has to be in order to meet the sufficient
condition. In other words, the Theorem indicates that the quality
of the learned controller $\hat{\kappa}$, in terms of low variability
of the control error, should be higher if the uncontrolled system
is more sensitive to input perturbations and the closed-loop system
obtained with $\kappa$ is closer to being unstable.\\
 The value of $\gamma_{\Delta}$ influences also the decay rate of
the term related to the initial condition $\|y(0)\|_{\infty}$, compare
eq. (\ref{E:stability_kappa_hat}), as well as the gain in the additive
term $\delta$. As to the latter, a comparison with the analogous
term in (\ref{E:stability_kappa}) reveals the effects of the absolute
value of the control error and of the presence of output noise. The
former is represented by the quantity $|\Delta_{0}|$, i.e. the magnitude
of the control error evaluated at $y=0$. Note that the choice of
$y=0$ to evaluate this term is not restrictive, since a simple coordinate
change can be used to refer all the results to a different output
value. According to the result, the is smaller the value of $|\Delta_{0}|$,
the closer is the term $\delta$ to the one obtained with the unknown
controller $\kappa$. This aspect, coupled with the condition (\ref{E:stability_cond})
on the value of $\gamma_{\Delta}$, basically states that, in order
to better replicate the behavior obtained with the controller $\kappa$,
the control error function has to be small in absolute value and have
low variability, as the intuition would suggest. About the noise term,
it contributes to $\delta$ in (\ref{E:stability_kappa_hat}) in a
way proportional to its maximum norm $\varepsilon_{y}$, and the gain
depends on how sensitive the controller $\hat{\kappa}$ is to perturbations
of its input argument, as indicated by the value of $\gamma_{\hat{\kappa}}$.
Finally, note that the effects of $|\Delta_{0}|$ and $\varepsilon_{y}$
are proportional to $\gamma_{f}$, i.e. to how sensitive the uncontrolled
plant is to input perturbations, and inversely proportional to $1-\gamma$,
i.e. to how close the closed loop system is to being unstable in the
sense of Definition \ref{D:stability}.

We extend next the stability analysis to the deviation between the
output trajectory obtained by using the controller $\kappa$ and the
one obtained by using $\hat{\kappa}$. In order to do so, we rename
as $\hat{y}(t)$ the output trajectory of system $\hat{g}$ (\ref{E:closed_loop_approx}),
and we define the deviation $\xi(t)\doteq\hat{y}(t)-y(t)$, where
$y$ is the output trajectory of the system $g$ defined in (\ref{E:closed_loop}).
Then, let us consider the following dynamical system:
\begin{equation}
\begin{array}{l}
\xi(t+1)=g_{\Delta}(\xi(t),y(t),e_{s}(t),e_{y}(t))\\
\;\doteq f(y(t)+\xi(t),\hat{\kappa}(y(t)+\xi(t)+e_{y}(t)),e_{s}(t))-f(y(t),\kappa(y(t)),e_{s}(t))\\
\;=f(\hat{y}(t),\hat{\kappa}(\hat{y}(t)+e_{y}(t)),e_{s}(t))-f(y(t),\kappa(y(t)),e_{s}(t))\\
g_{\Delta}:\Xi\times Y\times E\times B_{\varepsilon_{y}}\rightarrow Y,
\end{array}\label{E:closed_loop_dist}
\end{equation}
where $\Xi\subset\R^{n_{y}}$ is a compact set containing the values
of $\xi$, which is guaranteed to exist if the assumptions of Theorem
\ref{T:stability} hold, thanks to the combination of (\ref{E:stability_kappa})
and (\ref{E:stability_kappa_hat}).

\begin{corollary}\label{C:deviation_stab} Let Assumptions \ref{A:system_cont}-\ref{A:controller_stab}
and \ref{A:noise}-(b) hold. If (\ref{E:stability_cond}) holds, then
the system $g_{\Delta}$ is finite-gain $\ell_{\infty}$ stable. More
precisely, it holds that
\begin{equation}
\begin{array}{l}
\forall t\geq0,\\
\|\xi(t)\|_{\infty}\leq\underbrace{\dfrac{\gamma_{g,e}}{1-\gamma}\|\textbf{e}_{s}\|_{\infty}}\limits _{\alpha(\|\textbf{e}_{s}\|_{\infty})}+\underbrace{\gamma^{t}\|\xi(0)\|_{\infty}}\limits _{\beta(\|\xi(0)\|_{\infty},t)}+\underbrace{\dfrac{\gamma_{f}\gamma_{\Delta}}{1-\gamma}\gamma_{g}^{t}\|y(0)\|_{\infty}}\limits _{\beta_{y}(\|y(0)\|_{\infty},t)}+\underbrace{\dfrac{1}{1-\gamma}\left(\|g_{0}\|_{\infty}+\gamma_{f}|\Delta_{0}|+\gamma_{f}\gamma_{\hat{\kappa}}\varepsilon_{y}\right)}\limits _{\delta}.
\end{array}\label{E:stability_kappa_hat2}
\end{equation}
\end{corollary} \emph{Proof.} See the Appendix.\hfill{}$\square$

A comparison between the results (\ref{E:stability_kappa_hat}) and
(\ref{E:stability_kappa_hat2}) shows that the trajectory deviation
$\xi$ enjoys a closed-loop behavior, in the sense of Definition \ref{D:stability},
similar to the output of the closed loop system obtained with the
learned controller $\hat{\kappa}$, as far as the effects of $e_{s}(t)$,
$e_{y}(t)$ and $\Delta_{0}$ are concerned. The main difference with
respect to (\ref{E:stability_kappa_hat}) is the presence of a second
exponentially decaying term given by the function $\beta_{y}\in\mathcal{KL}$,
which depends on the initial condition $\|y(0)\|_{\infty}$. This
term can be interpreted as the relative effect of the magnitude of
the initial output on the magnitude of the trajectory deviation. The
practical meaning of the dependence of $\beta_{y}$ on the Lipschitz
constants $\gamma_{f}$, $\gamma_{\Delta}$, $\gamma_{g}$ and $\gamma$
can be deduced along the same lines of the comments on Theorem \ref{T:stability}.

The results presented so far serve as a theoretical justification
of the learning algorithm that we present in the next section, which
indeed is able to satisfy condition (\ref{E:stability_cond}) in the
limit, hence providing a solution to \textbf{Problem} 1.

\subsection{Learning algorithm}

\label{SS:approach}

A parametric representation is considered for the controller $\hat{\kappa}$:
\begin{equation}
\hat{\kappa}\left(y\right)=\sum_{i=1}^{M}\hat{a}_{i}\varphi_{i}\left(y\right)\label{f_star}
\end{equation}
where $\varphi_{i}:Y\rightarrow U$ are Lipschitz continuous basis
functions. The coefficients $\hat{a}_{i}\in\R$ are identified by
means of the following Algorithm \ref{des_alg}.

\begin{algorithm1}\label{des_alg}Controller learning.
\begin{enumerate}
\item \label{enu:prel_1}Take a set of basis functions $\left\{ \varphi_{i}\right\} _{i=1}^{M}$.
The choice of this set can be carried out by means of Procedure \ref{bf_ch}
below.
\item \label{enu:prel_2}Using the data set $\mathcal{D}^{N}\doteq\left\{ \tilde{u}(k),\tilde{y}(k)\right\} _{k=0}^{N-1}$
and the basis functions chosen at step \ref{enu:prel_1}), define
the following quantities:
\[
\begin{array}{l}
\Phi\doteq\left[\begin{array}{ccc}
\varphi_{1}\left(\widetilde{y}(0)\right) & \cdots & \varphi_{M}\left(\widetilde{y}(0)\right)\\
\vdots & \ddots & \vdots\\
\varphi_{1}\left(\widetilde{y}(N-1)\right) & \cdots & \varphi_{M}\left(\widetilde{y}(N-1)\right)
\end{array}\right]\in\R^{N\times M}\\
\widetilde{\boldsymbol{u}}\doteq(\widetilde{u}(0),\ldots,\widetilde{u}(N-1))\in\R^{N\times1}.
\end{array}
\]

\item \label{st_gg}Using Algorithms \ref{est_eps} and \ref{est_ga} below,
obtain an estimate $\hat{\varepsilon}$ of the noise bound $\varepsilon$
in (\ref{eq:zi_bound}), and estimates $\hat{\gamma}_{f}$ and $\hat{\gamma}_{g,y}$
of the Lipschitz constants $\gamma_{f}$ and $\gamma_{g,y}$ in (\ref{E:unif_cont_f})
and (\ref{E:unif_cont_y}). Choose $\gamma'_{\Delta}\simeq\left(1-\hat{\gamma}_{g,y}\right)/\hat{\gamma}_{f}$
such that $\gamma'_{\Delta}<\left(1-\hat{\gamma}_{g,y}\right)/\hat{\gamma}_{f}$.
\item \label{st_1}Solve the following convex optimization problem:
\begin{equation}
\begin{array}{l}
a^{1}=\arg\min\limits _{a\in\R^{M}}\left\Vert a\right\Vert _{1}\\
\text{subject to}\\
(a)\ \left\Vert \widetilde{\boldsymbol{u}}-\Phi a\right\Vert _{\infty}\leq\alpha\hat{\varepsilon}\\
(b)\ \left\vert \widetilde{u}(l)-\widetilde{u}(k)+\left(\Phi_{k}^{r}-\Phi_{l}^{r}\right)a\right\vert \leq\gamma'_{\Delta}\left\Vert \widetilde{y}(l)-\widetilde{y}(k)\right\Vert _{\infty}+2\hat{\varepsilon},\,\left\{ \begin{array}{l}
l=0,\ldots,N-1\\
k=l+1,\ldots,N-1
\end{array}\right.
\end{array}\label{opt21a}
\end{equation}
where $\Phi_{k}^{r}\doteq[\begin{array}{ccc}
\varphi_{1}\left(\widetilde{y}(k)\right) & \cdots & \varphi_{M}\left(\widetilde{y}(k)\right)\end{array}]$ and $\alpha\geq1$ is a number slightly larger than the minimum value
for which the constraint $(a)$ is feasible.
\item \label{st_2}Obtain the coefficient vector $\hat{a}=(\hat{a}_{1},\ldots,\hat{a}_{M})$
from the following convex optimization problem:
\begin{equation}
\begin{array}{l}
(\hat{a},\gamma_{\Delta}^{s})=\arg\min\limits _{a\in\R^{M},\,\gamma''_{\Delta}\in\R^{+}}\gamma''_{\Delta}\\
\text{subject to}\\
(a)\ \left\Vert \widetilde{\boldsymbol{u}}-\Phi a\right\Vert _{\infty}\leq\alpha\hat{\varepsilon}\\
(b)\ \left\vert \widetilde{u}(l)-\widetilde{u}(k)+\left(\Phi_{k}^{r}-\Phi_{l}^{r}\right)a\right\vert \leq\gamma''_{\Delta}\left\Vert \widetilde{y}(l)-\widetilde{y}(k)\right\Vert _{\infty}+2\hat{\varepsilon},\,\left\{ \begin{array}{l}
l=0,\ldots,N-1\\
k=l+1,\ldots,N-1
\end{array}\right.\\
(c)\ a_{i}=0,\ \forall i\notin\textrm{supp}\left(a^{1}\right)
\end{array}\label{opt21b}
\end{equation}
where $\textrm{supp}\left(a^{1}\right)$ is the support of $a^{1}$,
i.e. the set of indices at which $a^{1}$ is not null.\hfill{}$\square$
\end{enumerate}
\end{algorithm1}

The rationale behind the algorithm can be explained as follows. After
the preliminary operations carried out in steps \ref{enu:prel_1}-\ref{st_gg},
the $\ell_{1}$ norm of the coefficient vector $a$ is minimized in
step \ref{st_1}), leading to a sparse coefficient vector $a^{1}$,
i.e. a vector with a ``small'' number of non-zero elements. Constraint
$(a)$ in \eqref{opt21a} ensures the consistency between the measured
data and the prior information on the noise affecting these data (assuming
that $\hat{\varepsilon}$ is a reliable estimate of $\varepsilon$
and $\alpha$ is close to 1). Constraints $(b)$ allow us to guarantee
closed-loop stability when a sufficiently large number of data is
used, see Theorem \ref{lip_conv} below. Step \ref{st_2}) aims at
reducing the Lipschitz constant of the error function, maintaining
the same sparsity level obtained in step \ref{st_1}), and satisfying
the constraints for closed-loop stability. Indeed, the magnitude of
this constant is linked to the maximal deviation from the trajectory
achieved by the unknown controller $\kappa$, see Corollary \ref{C:deviation_stab},
hence step \ref{st_2}) of the algorithm accounts for the requirement
2) of \textbf{Problem} 1.\\
 The reason why a sparse controller is looked for is twofold. First,
a sparse function is easy to implement on real-time processors, which
may have limited memory and computational capacity, hence accounting
for the requirement 3) of \textbf{Problem} 1. Second, sparse functions
have nice regularity properties and are thus able to provide good
accuracy on new data by limiting well-known issues such as over-fitting
and the curse of dimensionality. A sparse function is a linear combination
of many basis functions, where the vector of linear combination coefficients
is sparse, i.e. it has only a few non-zero elements. The sparsity
of a vector is typically measured by the $\ell_{0}$ quasi-norm, defined
as the number of its non-zero elements. Sparse identification can
thus be performed by looking for a coefficient vector with a ``small''
$\ell_{0}$ quasi-norm. However, the $\ell_{0}$ quasi-norm is a non-convex
function and its minimization is in general an NP-hard problem. Two
main approaches are commonly adopted to deal with this issue: convex
relaxation and greedy algorithms \cite{Tropp04}, \cite{Fuchs05},
\cite{Tropp06}, \cite{Donoho06_2}. In convex relaxation, a suitable
convex function, e.g. the $\ell_{1}$ norm, is minimized instead of
the $\ell_{0}$ quasi-norm \cite{Fuchs05}, \cite{Tropp06}, \cite{Donoho06_2}.
In greedy algorithms, the sparse solution is obtained iteratively,
\cite{Tropp04}. Algorithm \ref{des_alg} is essentially an improved
$\ell_{1}$ algorithm: in step \ref{st_1}), an optimization problem
is solved, where the $\ell_{0}$ quasi-norm is replaced by the $\ell_{1}$
norm, and additional constraints for closed-loop stability are used
(i.e. $(b)$ in \eqref{opt21a}). In step \ref{st_2}), a vector $\hat{a}$
is obtained, with the same support as $a^{1}$, which minimizes the
estimated Lipschitz constant of the error function and satisfies the
closed-loop stability condition evaluated on the available data.

If a small number of data is used for control design, it may happen
that $\left(1-\hat{\gamma}_{g,y}\right)/\hat{\gamma}_{f}$$\leq0$,
thus not allowing a feasible choice of the Lipschitz constant $\gamma'_{\Delta}$
in step \ref{st_gg} of Algorithm \ref{des_alg}. In this case, our
indication is to collect a larger number of data in order to let the
estimated Lipschitz constants $\hat{\gamma}_{f}$ and $\hat{\gamma}_{g,y}$
get closer to the true ones, which by assumption satisfy the condition
$\left(1-\gamma_{g,y}\right)/\gamma_{f}$$\leq0$. Whether collecting
more data is not possible, our indication is to choose $\gamma'_{\Delta}$
slightly larger than the minimum value for which the optimization
problem (\ref{opt21a}) is feasible. Similar indications hold for
the case where $\left(1-\hat{\gamma}_{g,y}\right)/\hat{\gamma}_{f}>0$
but the chosen $\gamma'_{\Delta}$ is too small and constraint $(b)$
in (\ref{opt21a}) is thus not feasible.

\subsection{Parameter estimation and basis function choice}

\label{sub:par_bf}

\textcolor{black}{All the parameters involved in Algorithm \ref{des_alg}
(i.e. the noise bound $\hat{\varepsilon}$ and the Lipschitz constants
}$\hat{\gamma}_{f}$ and $\hat{\gamma}_{g,y}$\textcolor{black}{)
can be estimated in a systematic way }by means of the following Algorithms.

Suppose that a set of data $\{\widetilde{w}(k),\widetilde{z}(k)\}_{k=0}^{N-1}$
is available, described by
\begin{equation}
\widetilde{z}(k)=\mathfrak{f}\left(\widetilde{w}(k)\right)+e(k),\: k=0,\ldots,N-1\label{eq:meas_2}
\end{equation}
where $\mathfrak{f}:W\rightarrow\mathbb{R}$ is a generic unknown
function, $W\subset\mathbb{R}^{n_{w}}$ and $e(k)$ is an unknown
noise. Assume that $e(k)\in B_{\varepsilon},\:\forall k,$ and $\mathfrak{f}$
is Lipschitz continuous with constant $\gamma_{\mathfrak{f}}$. The
noise bound $\varepsilon$ and the Lipschitz constant $\gamma_{\mathfrak{f}}$
can be estimated as follows.

\begin{algorithm1}\label{est_eps}Noise bound estimation.
\begin{enumerate}
\item \label{enu:rho}Choose a ``small'' $\rho>0$. For example: $\rho=0.01\max\limits _{k,l=0,\ldots,N-1}\left\Vert \widetilde{w}(k)-\widetilde{w}(l)\right\Vert _{\infty}$.
\item \label{enu:dz}For $k=0,\ldots,N-1$, compute
\[
\delta\tilde{z}_{k}=\max_{i,j\in J_{k}}\left|\widetilde{z}(i)-\widetilde{z}(j)\right|
\]
where $J_{k}\doteq\left\{ l:\left\Vert \widetilde{w}(k)-\widetilde{w}(l)\right\Vert _{\infty}\leq\rho\right\} $.
If $J_{k}=\emptyset$, set $\delta\tilde{z}_{k}=\infty$. If $J_{k}=\emptyset$
for all $k=0,\ldots,N-1$ , go to step \ref{enu:rho}) and choose
a larger $\rho$.
\item \label{enu:eps_hat}Obtain the estimate $\hat{\varepsilon}$ of the
noise bound $\varepsilon$ as
\[
\hat{\varepsilon}=\frac{1}{2\hat{N}}\sum_{k\in Q}\delta\tilde{z}_{k}
\]
where $Q\doteq\left\{ k\in\left\{ 0,\ldots,N-1\right\} :\delta\tilde{z}_{k}<\infty\right\} $
and $\hat{N}\doteq\textrm{card}\left(Q\right)$.
\end{enumerate}
\hfill{}$\square$

\end{algorithm1}

\begin{algorithm1}\label{est_ga}Lipschitz constant estimation.
\begin{enumerate}
\item For $k,l=0,\ldots,N-1$ and $\widetilde{w}(k)\neq\widetilde{w}(l)$,
compute
\[
\tilde{\gamma}_{lk}=\left\{ \begin{array}{cl}
\frac{\widetilde{z}(k)-\widetilde{z}(l)-2\hat{\varepsilon}}{\left\Vert \widetilde{w}(k)-\widetilde{w}(l)\right\Vert _{\infty}}, & \textrm{if }\widetilde{z}(k)>\widetilde{z}(l)+2\hat{\varepsilon}\\
\frac{\widetilde{z}(l)-\widetilde{z}(k)-2\hat{\varepsilon}}{\left\Vert \widetilde{w}(k)-\widetilde{w}(l)\right\Vert _{\infty}}, & \textrm{if }\widetilde{z}(l)>\widetilde{z}(k)+2\hat{\varepsilon}\\
0, & \textrm{otherwise}
\end{array}\right.
\]
where $\hat{\varepsilon}$ is the noise bound estimated by Algorithm
\ref{est_eps}.
\item Obtain the estimate $\hat{\gamma}$ of the Lipschitz constant $\gamma_{\mathfrak{f}}$
as
\[
\hat{\gamma}=\max\limits _{k,l=0,\ldots,N-1:\widetilde{w}(k)\neq\widetilde{w}(l)}\tilde{\gamma}_{lk}
\]

\end{enumerate}
\hfill{}$\square$

\end{algorithm1}

The two algorithms above allow the estimation of the Lipschitz constant
of a generic function $\mathfrak{f}$. We now discuss how the algorithms
can be applied to estimate the Lipschitz constants of the functions
$f$ in \eqref{E:sys2} and $g$ in (\ref{E:closed_loop}), which
are required by the learning algorithm \ref{des_alg}. The Lipschitz
constant $\gamma_{f}$ of the function $f$ (with respect to $u(k)$)
can be estimated considering that
\[
\tilde{y}(k+1)=\mathfrak{f}(\tilde{u}(k))+v_{f}(k),\: k=0,\ldots,N-1
\]
where $\mathfrak{f}(\tilde{u}(t))\doteq f(y^{*},\tilde{u}(k),e_{s}^{*})$
is an unknown function with Lipschitz constant$\gamma_{f}$, the quantities
$y^{*}$ and $e_{s}^{*}$ are defined as
\[
\left(y^{*},e_{s}^{*}\right)=\arg\max_{\left(y,e\right)\in Y\times E_{s}}\mathfrak{L}_{f}\left(y,e\right),\:\mathfrak{L}_{f}\left(y,e\right)\doteq\max_{u^{1},u^{2}\in U}\frac{\left\Vert f(y,u^{1},e)-f(y,u^{2},e)\right\Vert _{\infty}}{\left|u^{1}-u^{2}\right|}
\]
and
\begin{equation}
v_{f}(k)\doteq f(y(k),u(k),e_{s}(k))-f(y^{*},\tilde{u}(k),e_{s}^{*})+e_{y}(k+1)\label{eq:vf_def}
\end{equation}
is an unknown noise. Analogously, the Lipschitz constant $\gamma_{g,y}$
of the function $g$ (with respect to $y(k)$) can be estimated considering
that
\[
\tilde{y}(k+1)=\mathfrak{g}(\tilde{y}(k))+v_{g}(k),\: k=0,\ldots,N-1
\]
where $\mathfrak{g}(\tilde{y}(t))\doteq g(\tilde{y}(k),e_{s}^{*})$
is an unknown function with \textcolor{black}{Lipschitz constant }$\gamma_{g,y}$,
the quantity $e_{s}^{*}$ is defined as
\[
e_{s}^{*}=\arg\max_{e\in E_{s}}\mathfrak{L}_{g}\left(e\right),\:\mathfrak{L}_{g}\left(e\right)\doteq\max_{y^{1},y^{2}\in Y}\frac{\left\Vert g(y^{1},e)-g(y^{2},e)\right\Vert _{\infty}}{\left|y^{1}-y^{2}\right|}
\]
and
\begin{equation}
v_{g}(k)\doteq g(y(k),e_{s}(k))-g(\tilde{y}(k),e_{s}^{*})+e_{y}(k+1)\label{eq:vg_def}
\end{equation}
is an unknown noise. Note that the noises $v_{f}(k)$ and $v_{g}(k)$
in (\ref{eq:vf_def}) and (\ref{eq:vg_def}) are bounded, due to the
boundedness of $u(k)$, $e_{s}(k)$ and $y(k)$ and the Lipschitz
continuity of $f$ and $g$: $v_{f}(k)\in B_{\varepsilon_{f}},\: v_{g}(k)\in B_{\varepsilon_{g}},\:\forall k\geq0$.

Another important step of Algorithm \ref{des_alg} is the choice of
the basis functions $\varphi_{i}$ (as well known, this aspect is
crucial for any identification method relying on a basis function
representation). An inappropriately chosen family of functions can
force the retention of many terms by the identification algorithm
or can lead to large approximation errors. In these situations, several
problems may arise, such as high controller complexity, closed-loop
instability and/or large deviations from the ideal trajectory. The
following procedure can be used to address this issue.

\begin{procedure}\label{bf_ch}Choice of basis function family.
\begin{enumerate}
\item \label{enu:bfc1}Run Algorithm \ref{des_alg} using a given family
of basis functions (e.g. Gaussian, sigmoidal, wavelet, polynomial,
trigonometric).
\item Consider the following cases:

\begin{enumerate}
\item If $\alpha$ is small (close to 1) and $\hat{a}$ is sparse, it may
be concluded that the basis functions have been correctly chosen,
since a small number of them is able to ``explain\textquotedblright{}
the measured data. Then, stop the procedure.
\item If $\alpha$ is small and $\hat{a}$ is not sparse, it may be guessed
that the basis function choice is not appropriate. Indeed, using a
large number of basis functions may lead to overfitting problems and
possibly to closed-loop instability. Then, go back to step \ref{enu:bfc1}),
choosing a different family of basis functions.
\item If $\alpha$ is not small, it may be guessed that the basis function
choice is not appropriate since it leads to a large approximation
error on the measured data. Then, go back to step \ref{enu:bfc1}),
choosing a different family of basis functions. \hfill{}$\square$
\end{enumerate}
\end{enumerate}
\end{procedure} The quality of the derived approximation can be also
assessed by testing it on data that were not used in the learning
algorithm, as it is commonly done in identification problems.

\subsection{Asymptotic analysis}

In this subsection, the asymptotic properties of Algorithms \ref{des_alg},
\ref{est_eps} and \ref{est_ga} are analyzed. In particular, the
following theorems show that the noise bound estimate $\hat{\varepsilon}$
and the Lipschitz constant estimate $\hat{\gamma}$ provided by Algorithms
\ref{est_eps} and \ref{est_ga}, respectively, converge to the true
values when the number $N$ of data tends to infinity.

\begin{theorem} \label{thm:conv1}Let the set $\{\left(\widetilde{w}(k),e(k)\right)\}_{k=0}^{N-1}$
appearing in (\ref{eq:meas_2}) be dense on $W\times B_{\varepsilon}$
as $N\rightarrow\infty$. Then,
\[
\lim_{N\rightarrow\infty}\hat{\varepsilon}=\varepsilon
\]
where $\hat{\varepsilon}$ is the noise bound estimated by Algorithm
\ref{est_eps}.

\end{theorem}\emph{Proof.} See the Appendix.\hfill{}$\square$

\begin{theorem} \label{thm:conv2}Let the set $\{\left(\widetilde{w}(k),e(k)\right)\}_{k=0}^{N-1}$
appearing in (\ref{eq:meas_2}) be dense on $W\times B_{\varepsilon}$
as $N\rightarrow\infty$. Then,
\[
\lim_{N\rightarrow\infty}\hat{\gamma}=\gamma_{\mathfrak{f}}
\]
where $\hat{\gamma}$ is the Lipschitz constant estimated by Algorithm
\ref{est_ga}.

\end{theorem}\emph{Proof.} See the Appendix.\hfill{}$\square$

A result is now presented, showing that the controller $\hat{\kappa}$
identified by means of Algorithm \ref{des_alg} satisfies the stability
condition (\ref{E:stability_cond}) when the number of data $N$ tends
to infinity. Before stating the result, let us introduce two technical
assumptions, regarding the noises $v_{f}(k)$ and $v_{g}(k)$ defined
in (\ref{eq:vf_def}) and (\ref{eq:vg_def}), respectively.

\begin{assumption}\label{dense_ass-1}The set of points $\mathcal{D}_{uv}^{N}\doteq\left\{ \widetilde{u}(k),v_{f}(k)\right\} _{k=0}^{N-1}$
is dense on $U\times B_{\varepsilon_{f}}$ as $N\rightarrow\infty$.
\hfill{}$\square$

\end{assumption}

\begin{assumption}\label{dense_ass-2}The set of points $\mathcal{D}_{yv}^{N}\doteq\left\{ \left(\widetilde{y}(k),v_{g}(k)\right)\right\} _{k=0}^{N-1}$
is dense on $Y\times B_{\varepsilon_{g}}$ as $N\rightarrow\infty$.\hfill{}$\square$

\end{assumption} In Assumptions \ref{dense_ass-1}-\ref{dense_ass-2},
density of the sets $\mathcal{D}_{uv}^{N},\,\mathcal{D}_{yv}^{N}$
is intended in the same sense as in Assumption \ref{dense_ass}.

\begin{theorem} \label{lip_conv}Let the optimization problem (\ref{opt21a})
be feasible for any $N\geq0$. Let Assumptions \ref{A:system_cont}-\ref{A:controller_cont},
\ref{A:noise}, and \ref{dense_ass}-\ref{dense_ass-2} hold. Then,
the error function $\Delta\doteq\kappa-\hat{\kappa}$ is Lipschitz
continuous on $Y$, with constant $\gamma_{\Delta}$ such that
\[
\limsup_{N\rightarrow\infty}\gamma_{\Delta}\leq\gamma_{\Delta}^{s}<\dfrac{1-\gamma_{g,y}}{\gamma_{f}}.
\]
\end{theorem}

\emph{Proof.} See the Appendix.\hfill{}$\square$

We illustrate next the convergence result of Theorem \ref{lip_conv}
through a simple numerical example.\medskip{}

\textbf{Example: asymptotic behavior of the estimated Lipschitz constant.}\medskip{}

\begin{figure}
\centering

\includegraphics[scale=0.7]{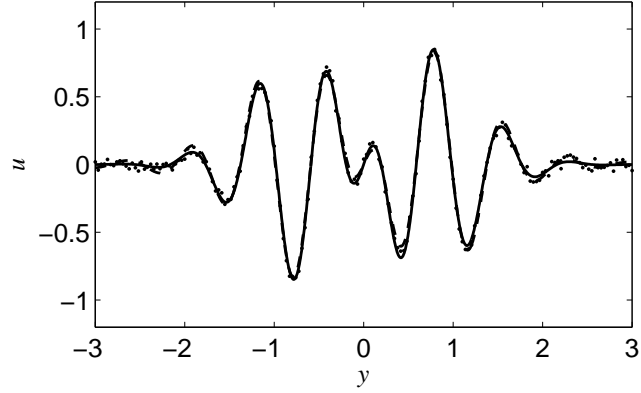}

\caption{Continuous: true function $\kappa$. Dashed: Algorithm \ref{des_alg}
estimate $\hat{\kappa}^{170}$. Dots: measurements.}

\label{Fig:lip_conv-2}
\end{figure}

We have considered the function
\[
u=\kappa\left(y\right)\doteq2ye^{-y^{2}}\cos\left(8y\right),
\]
shown in Fig. \ref{Fig:lip_conv-2}, and values of $N=10,20,\ldots,250$.
For each one of these values, we generated a data set $\mathcal{D}^{N}\doteq\left\{ \tilde{u}(k),\tilde{y}(k)\right\} _{k=0}^{N-1}$
according to
\[
\widetilde{u}(k)=\kappa\left(\widetilde{y}(k)\right)+d(k),\: k=0,\ldots,N-1
\]
where $d(k)$ is a white uniform noise with amplitude 0.05. Then,
we applied Algorithm \ref{des_alg} to obtain an estimate $\hat{\kappa}^{N}$
of $\kappa$ of the form (\ref{f_star}), where $\varphi_{i}:[-3,3]\rightarrow[-1,1]$
are Gaussian basis functions:
\[
\varphi_{i}(y)=e^{-100(y-\tilde{y}(i))^{2}},\: i=0,\ldots,N-1.
\]
For comparison, we computed another estimate $\hat{\kappa}_{nc}^{N}$
of the same form (\ref{f_star}), using the same Gaussian basis functions,
by means of Algorithm \ref{des_alg}, but without the constraints
$(b)$ in \eqref{opt21a}. We recall that, according to Theorem \ref{lip_conv},
these constraints yield the convergence of the Lipschitz constant
of the error function to a value that ensures closed-loop stability.

\begin{figure}
\centering

\includegraphics[scale=0.7]{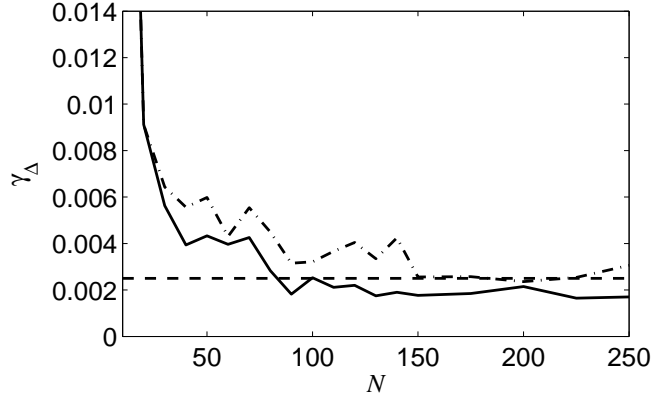}

\caption{Continuous: $\gamma_{\Delta}^{N}$. Dot-dashed: $\gamma_{\Delta nc}^{N}$.
Dashed: hypothetical stability bound.}

\label{Fig:lip_conv-1}
\end{figure}

The Lipschitz constant $\gamma_{\Delta}^{N}$ of the error function
$\Delta^{N}\doteq\kappa-\hat{\kappa}^{N}$ and the Lipschitz constant
$\gamma_{\Delta nc}^{N}$ of the error function $\Delta^{N}\doteq\kappa-\hat{\kappa}_{nc}^{N}$
are shown in Fig. \ref{Fig:lip_conv-1} for $N=10,\,20,\,\ldots,\,250$.
It can be noted that $\gamma_{\Delta}^{N}$ decreases quite rapidly
as $N$ becomes large, taking soon values below an hypothetical threshold
that is sufficient for closed-loop stability. Also $\gamma_{\Delta nc}^{N}$
seems to have a similar behavior, however the decrease is slower and
less regular with respect to the one of $\gamma_{\Delta}^{N}$ and,
in any case, satisfaction of the stability condition is not ensured
theoretically. In Fig. \ref{Fig:lip_conv-2}, the estimate $\hat{\kappa}^{170}$
is compared with the true function $\kappa$.

\section*{Appendix}

\noindent \emph{Derivation of eq. (\ref{E:stability_kappa})}. Consider
$t=0$. We have:
\[
\begin{array}{cl}
\|y(1)\|_{\infty} & =\|g(y(0),e_{s}(0))\|_{\infty}\\
 & =\|g(y(0),e_{s}(0))-g(0,e_{s}(0))+g(0,e_{s}(0))-g_{0}+g_{0}\|_{\infty}\\
 & \leq\|g(y(0),e_{s}(0))-g(0,e_{s}(0))\|_{\infty}+\|g(0,e_{s}(0))-g_{0}\|_{\infty}+\|g_{0}\|_{\infty}.
\end{array}
\]
Using properties \eqref{E:unif_cont_y}-\eqref{E:unif_cont_e}, we
obtain
\[
\begin{array}{lll}
\|y(1)\|_{\infty} & \leq & \gamma_{g,y}\|y(0)\|_{\infty}+\gamma_{g,e}\|e_{s}(0)\|_{\infty}+\|g_{0}\|_{\infty}\\
 & \leq & \gamma_{g,y}\|y(0)\|_{\infty}+\gamma_{g,e}\|\textbf{e}_{s}\|_{\infty}+\|g_{0}\|_{\infty}.
\end{array}
\]
Analogously,
\[
\begin{array}{lll}
\|y(2)\|_{\infty} & \leq & \gamma_{g,y}^{2}\|y(0)\|_{\infty}+\gamma_{g,y}\gamma_{g,e}\|\textbf{e}_{s}\|_{\infty}\\
 &  & +\gamma_{g,e}\|\textbf{e}_{s}\|_{\infty}+\gamma_{g,y}\|g_{0}\|_{\infty}+\|g_{0}\|_{\infty}.
\end{array}
\]
The result is then established by generalizing to any $t\geq0$:

\noindent
\[
\begin{array}{lll}
\|y(t)\|_{\infty} & \leq & \gamma_{g,y}^{t}\|y(0)\|_{\infty}+\sum\limits _{k=0}^{t-1}\left(\gamma_{g,y}^{k}\gamma_{g,e}\|\textbf{e}_{s}\|_{\infty}\right)+\sum\limits _{k=0}^{t-1}\left(\gamma_{g,y}^{k}\|g_{0}\|_{\infty}\right)\\
 & \leq & \dfrac{1}{1-\gamma_{g,y}}\gamma_{g,e}(\|\textbf{e}_{s}\|_{\infty})+\gamma_{g,y}^{t}\|y(0)\|_{\infty}+\dfrac{1}{1-\gamma_{g,y}}\|g_{0}\|_{\infty},\,\forall t\geq0.
\end{array}
\]

\emph{Proof of Theorem \ref{T:stability}}. Consider $t=0$. We have
\[
\begin{array}{l}
\|y(1)\|_{\infty}=\|\hat{g}(y(0),e_{s}(0),e_{y}(0))\|_{\infty}\\
=\|\hat{g}(y(0),e_{s}(0),e_{y}(0))-g(y(0),e_{s}(0))+g(y(0),e_{s}(0))\|_{\infty}.
\end{array}
\]
Using properties \eqref{E:unif_cont_f} and \eqref{E:unif_cont_y}-\eqref{E:unif_cont_e},
\[
\begin{array}{l}
\|\hat{g}(y(0),e_{s}(0),e_{y}(0))-g(y(0),e_{s}(0))+g(y(0),e_{s}(0))\|_{\infty}\\
\leq\|\hat{g}(y(0),e_{s}(0),e_{y}(0))-g(y(0),e_{s}(0))\|_{\infty}+\|g(y(0),e_{s}(0))\|_{\infty}\\
=\|f(y(0),\hat{\kappa}(y(0)+e_{y}(0)),e_{s}(0))-f(y(0),\kappa(y(0)),e_{s}(0))\|_{\infty}\\
+\|g(y(0),(0))-g(0,0)+g_{0}\|_{\infty}\\
\leq\gamma_{f}|\hat{\kappa}(y(0)+e_{y}(0))-\kappa(y(0))|+\gamma_{g,y}\|y(0)\|_{\infty}+\gamma_{g,e}\|e_{s}(0)\|_{\infty}+\|g_{0}\|_{\infty}
\end{array}
\]
where we recall that $g(0,0)\doteq g_{0}$. Using properties \eqref{E:approx_lipschitz}
and \eqref{E:error_lipschitz},
\[
\begin{array}{l}
\gamma_{f}|\hat{\kappa}(y(0)+e_{y}(0))-\kappa(y(0))|+\gamma_{g,y}\|y(0)\|_{\infty}+\gamma_{g,e}\|e_{s}(0)\|_{\infty}+\|g_{0}\|_{\infty}\\
=\gamma_{f}|\hat{\kappa}(y(0)+e_{y}(0)))-\hat{\kappa}(y(0))+\hat{\kappa}(y(0))-\kappa(y(0))|+\gamma_{g,y}\|y(0)\|_{\infty}+\gamma_{g,e}\|e_{s}(0)\|_{\infty}+\|g_{0}\|_{\infty}\\
=\gamma_{f}|\Delta(y(0))-\Delta(0)+\Delta_{0}|+\gamma_{f}\gamma_{\hat{\kappa}}\|e_{y}(0)\|_{\infty}+\gamma_{g,y}\|y(0)\|_{\infty}+\gamma_{g,e}\|e_{s}(0)\|_{\infty}+\|g_{0}\|_{\infty}\\
\leq\gamma_{f}\gamma_{\Delta}\|y(0)\|_{\infty}+\gamma_{f}|\Delta_{0}|+\gamma_{f}\gamma_{\hat{\kappa}}\|e_{y}(0)\|_{\infty}+\gamma_{g,y}\|y(0)\|_{\infty}+\gamma_{g,e}\|\textbf{e}_{s}\|_{\infty}+\|g_{0}\|_{\infty}\\
\leq(\gamma_{f}\gamma_{\Delta}+\gamma_{g,y})\|y(0)\|_{\infty}+\gamma_{g,e}\|\textbf{e}_{s}\|_{\infty}+\|g_{0}\|_{\infty}+\gamma_{f}|\Delta_{0}|+\gamma_{f}\gamma_{\hat{\kappa}}\varepsilon_{y}\\
=\gamma\|y(0)\|_{\infty}+\gamma_{g,e}\|\textbf{e}_{s}\|_{\infty}+\|g_{0}\|_{\infty}+\gamma_{f}|\Delta_{0}|+\gamma_{f}\gamma_{\hat{\kappa}}\varepsilon_{y}
\end{array}
\]
where we recall that $\Delta(0,0)\doteq\Delta_{0}$ and $\gamma\doteq(\gamma_{f}\gamma_{\Delta}+\gamma_{g,y})<1$.
Analogously, for $t=1$, we have that
\[
\begin{array}{l}
\|y(2)\|_{\infty}\leq\gamma\|y(1)\|_{\infty}+\gamma_{g,e}\|\textbf{e}_{s}\|_{\infty}+\|g_{0}\|_{\infty}+\gamma_{f}|\Delta_{0}|+\gamma_{f}\gamma_{\hat{\kappa}}\varepsilon_{y}\\
\leq\gamma^{2}\|y(0)\|_{\infty}+\gamma\gamma_{g,e}\|\textbf{e}_{s}\|_{\infty}+\gamma\left(\|g_{0}\|_{\infty}+\gamma_{f}|\Delta_{0}|+\gamma_{f}\gamma_{\hat{\kappa}}\varepsilon_{y}\right)\\
+\gamma_{g,e}\|\textbf{e}_{s}\|_{\infty}+\|g_{0}\|_{\infty}+\gamma_{f}|\Delta_{0}|+\gamma_{f}\gamma_{\hat{\kappa}}\varepsilon_{y}.
\end{array}
\]
Generalizing to any $t\geq0$, we obtain
\[
\begin{array}{c}
\|y(t)\|_{\infty}\leq\gamma^{t}\|y(0)\|_{\infty}+\sum\limits _{k=0}^{t-1}\gamma^{k}\gamma_{g,e}\|\textbf{e}_{s}\|_{\infty}+\sum\limits _{k=0}^{t-1}\left(\gamma^{k}\left(\|g_{0}\|_{\infty}+\gamma_{f}|\Delta_{0}|+\gamma_{f}\gamma_{\hat{\kappa}}\varepsilon_{y}\right)\right).\end{array}
\]
Considering \eqref{E:stability_cond} and the convergence of the geometric
series, it follows that
\[
\begin{array}{c}
\|y(t)\|_{\infty}\leq\dfrac{\gamma_{g,e}}{1-\gamma}(\|\textbf{e}_{s}\|_{\infty})+\gamma^{t}\|y(0)\|_{\infty}+\dfrac{1}{1-\gamma}\left(\|g_{0}\|_{\infty}+\gamma_{f}|\Delta_{0}|+\gamma_{f}\gamma_{\hat{\kappa}}\varepsilon_{y}\right),\,\forall t\geq0,\end{array}
\]
which proves the claim.

\emph{Proof of Corollary \ref{C:deviation_stab}}. Consider a generic
time $t\geq0$. We have
\[
\begin{array}{l}
\|\xi(t+1)\|_{\infty}=\|f(\hat{y}(t),\hat{\kappa}(\hat{y}(t)+e_{y}(t)),e_{s}(t))-f(y(t),\kappa(y(t)),e_{s}(t))\|_{\infty}\\
=\left\Vert f(\hat{y}(t),\hat{\kappa}(\hat{y}(t)+e_{y}(t)),e_{s}(t))-f(\hat{y}(t),\kappa(\hat{y}(t)),e_{s}(t))\right.\\
+\left.f(\hat{y}(t),\kappa(\hat{y}(t)),e_{s}(t))-f(y(t),\kappa(y(t)),e_{s}(t))\right\Vert _{\infty}\\
=\left\Vert f(\hat{y}(t),\hat{\kappa}(\hat{y}(t)+e_{y}(t)),e_{s}(t))-f(\hat{y}(t),\kappa(\hat{y}(t)),e_{s}(t))\right.\\
+\left.g(\hat{y}(t),e_{s}(t))-g(y(t),e_{s}(t))\right\Vert _{\infty}\\
\leq\gamma_{f}|\hat{\kappa}(\hat{y}(t)+e_{y}(t))-\kappa(\hat{y}(t))|+\gamma_{g,y}\|\xi(t)\|_{\infty}
\end{array}
\]
where properties \eqref{E:unif_cont_f} and \eqref{E:unif_cont_y}
have been used in the last inequality. Moreover,
\[
\begin{array}{l}
\gamma_{f}|\hat{\kappa}(\hat{y}(t)+e_{y}(t))-\kappa(\hat{y}(t))|+\gamma_{g,y}\|\xi(t)\|_{\infty}\\
=\gamma_{f}|\hat{\kappa}(\hat{y}(t)+e_{y}(t))-\hat{\kappa}(\hat{y}(t))+\hat{\kappa}(\hat{y}(t))-\kappa(\hat{y}(t))|+\gamma_{g,y}\|\xi(t)\|_{\infty}\\
=\gamma_{f}|\Delta(\hat{y}(t))-\Delta_{0}+\Delta_{0}+\hat{\kappa}(\hat{y}(t)+e_{y}(t))-\hat{\kappa}(\hat{y}(t))|+\gamma_{g,y}\|\xi(t)\|_{\infty}.
\end{array}
\]
Using properties \eqref{E:approx_lipschitz}, \eqref{E:error_lipschitz},
and the equality $\hat{y}(t)=\xi(t)+y(t)$, we obtain that
\[
\begin{array}{c}
\gamma_{f}|\Delta(\hat{y}(t))-\Delta_{0}+\Delta_{0}+\hat{\kappa}(\hat{y}(t)+e_{y}(t))-\hat{\kappa}(\hat{y}(t))|+\gamma_{g,y}\|\xi(t)\|_{\infty}\\
\leq\gamma_{f}\gamma_{\Delta}\|\xi(t)\|_{\infty}+\gamma_{f}\gamma_{\Delta}\|y(t)\|_{\infty}+\gamma_{f}|\Delta_{0}|+\gamma_{f}\gamma_{\hat{\kappa}}\|e_{y}(t)\|_{\infty}+\gamma_{g,y}\|\xi(t)\|_{\infty}\\
=\gamma\|\xi(t)\|_{\infty}+\gamma_{f}\gamma_{\Delta}\|y(t)\|_{\infty}+\gamma_{f}|\Delta_{0}|+\gamma_{f}\gamma_{\hat{\kappa}}\|e_{y}(t)\|_{\infty}
\end{array}
\]
where $\gamma\doteq(\gamma_{f}\gamma_{\Delta}+\gamma_{g,y})<1$. Using
\eqref{E:stability_kappa}, iterating backwards to $t=0$ and considering
inequality \eqref{E:stability_cond} and the convergence of the geometric
series,
\[
\begin{array}{l}
\|\xi(t+1)\|_{\infty}\leq\gamma^{t}\|\xi(0)\|_{\infty}+\dfrac{\gamma_{f}}{1-\gamma}(\gamma_{\hat{\kappa}}\varepsilon_{y}+|\Delta_{0}|)\\
+\dfrac{\gamma_{f}\gamma_{\Delta}}{1-\gamma}\left(\dfrac{\gamma_{g,e}}{1-\gamma_{g,y}}\|\textbf{e}_{s}\|_{\infty}+\gamma_{g,y}^{t}\|y(0)\|_{\infty}+\dfrac{1}{1-\gamma_{g,y}}\|g_{0}\|_{\infty}\right),
\end{array}
\]
which establishes the result.

\emph{Proof of Theorem \ref{thm:conv1}}. Consider step \ref{enu:dz})
of Algorithm \ref{est_eps}. Equations (\ref{eq:meas_2}) imply that
\[
\begin{array}{l}
\delta\tilde{z}_{k}=\max\limits _{i,j\in J_{k}}\left|\widetilde{z}(i)-\widetilde{z}(j)\right|\\
=\max\limits _{i,j\in J_{k}}\left|\mathfrak{f}\left(\widetilde{w}(i)\right)-\mathfrak{f}\left(\widetilde{w}(j)\right)+e(i)-e(j)\right|\\
\geq\max\limits _{i,j\in J_{k}}\left(\left|e(i)-e(j)\right|-\left|\mathfrak{f}\left(\widetilde{w}(i)\right)-\mathfrak{f}\left(\widetilde{w}(j)\right)\right|\right).
\end{array}
\]
From Assumption \ref{dense_ass}, it follows that, for any $\lambda>0$,
there exist a sufficiently large $N$ and two pairs $\left(\widetilde{w}(i),e(i)\right)\in\left\{ \widetilde{w}(i)\right\} _{k=0}^{N-1}\times B_{e}$
and $\left(\widetilde{w}(j),e(j)\right)\in\left\{ \widetilde{w}(i)\right\} _{k=0}^{N-1}\times B_{e}$
with $i,j\in J_{k}$ such that
\[
\left|\varepsilon-e(i)\right|\leq\lambda,\:\left|-\varepsilon-e(j)\right|\leq\lambda,
\]
thus yielding the following inequality:
\[
\left|e(i)-e(j)\right|\geq2\varepsilon-2\lambda.
\]
Moreover, due the Lipschitz continuity property, we have
\[
\left|\mathfrak{f}\left(\widetilde{w}(i)\right)-\mathfrak{f}\left(\widetilde{w}(j)\right)\right|\leq\gamma_{\mathfrak{f}}\left\Vert \widetilde{w}(i)-\widetilde{w}(j)\right\Vert _{\infty}\leq2\gamma_{\mathfrak{f}}\rho.
\]
The above inequalities imply that
\begin{equation}
\begin{array}{l}
\delta\tilde{z}_{k}\geq\max\limits _{i,j\in J_{k}}\left(\left|e(i)-e(j)\right|-\left|\mathfrak{f}\left(\widetilde{w}(i)\right)-\mathfrak{f}\left(\widetilde{w}(j)\right)\right|\right)\\
\geq\max\limits _{i,j\in J_{k}}\left(\left(\left|e(i)-e(j)\right|\right)-2\gamma_{\mathfrak{f}}\rho\right)\\
\geq2\varepsilon-2\lambda-2\gamma_{\mathfrak{f}}\rho.
\end{array}\label{eq:c1_pr_1}
\end{equation}
On the other hand,
\begin{equation}
\begin{array}{l}
\delta\tilde{z}_{k}=\max\limits _{i,j\in J_{k}}\left|\widetilde{z}(i)-\widetilde{z}(j)\right|\\
=\max\limits _{i,j\in J_{k}}\left|\mathfrak{f}\left(\widetilde{w}(i)\right)-\mathfrak{f}\left(\widetilde{w}(j)\right)+e(i)-e(j)\right|\\
\leq\max\limits _{i,j\in J_{k}}\left(\left|e(i)-e(j)\right|+\left|\mathfrak{f}\left(\widetilde{w}(i)\right)-\mathfrak{f}\left(\widetilde{w}(j)\right)\right|\right)\\
\leq2\varepsilon+2\gamma_{\mathfrak{f}}\rho.
\end{array}\label{eq:c1_pr_2}
\end{equation}
Since $\lambda$ and $\rho$ can be chosen arbitrarily small, from
(\ref{eq:c1_pr_1}) and (\ref{eq:c1_pr_2}) it follows that $\delta\tilde{z}_{k}\rightarrow2\varepsilon$
as $N\rightarrow\infty$, i.e. that $\delta\tilde{z}_{k}/2\rightarrow\varepsilon$.
In step \ref{enu:eps_hat}) of Algorithm \ref{est_eps}, the operation
of taking the mean over all $\delta\tilde{z}_{k}/2$ is inessential
in this asymptotic analysis. It can be effective in the finite data
case in order to not under-estimate or over-estimate $\varepsilon$.

\emph{Proof of Theorem \ref{thm:conv2}}. Define
\[
(\bar{w}^{1},\bar{w}^{2})\doteq\arg\max_{w^{1},w^{2}\in W}\frac{\left|\mathfrak{f}(w^{1})-\mathfrak{f}(w^{2})\right|}{\left\Vert w^{1}-w^{2}\right\Vert _{\infty}}
\]
and, without loss of generality, suppose that $\mathfrak{f}(\bar{w}^{1})>\mathfrak{f}(\bar{w}^{2})$.
From Assumption \ref{dense_ass}, it follows that, for any $\lambda>0$,
there exist a sufficiently large $N$ and two pairs $\left(\widetilde{w}(i),e(i)\right)\in\left\{ \widetilde{w}(i)\right\} _{k=0}^{N-1}\times B_{e}$
and $\left(\widetilde{w}(j),e(j)\right)\in\left\{ \widetilde{w}(i)\right\} _{k=0}^{N-1}\times B_{e}$
with $i,j\in\{0,\ldots N-1\}$ such that
\begin{equation}
\begin{array}{c}
\left\Vert \bar{w}^{1}-\widetilde{w}(i)\right\Vert _{\infty}\leq\lambda,\:\left\Vert \bar{w}^{2}-\widetilde{w}(j)\right\Vert _{\infty}\leq\lambda\\
\left|\varepsilon-e(i)\right|\leq\lambda,\:\left|-\varepsilon-e(j)\right|\leq\lambda.
\end{array}\label{eq:cv2_pr_1}
\end{equation}
Moreover,
\begin{equation}
\begin{array}{c}
\left\Vert \bar{w}^{1}-\bar{w}^{2}\right\Vert _{\infty}=\left\Vert \bar{w}^{1}-\widetilde{w}(i)-\bar{w}^{2}+\widetilde{w}(j)+\widetilde{w}(i)-\widetilde{w}(j)\right\Vert _{\infty}\\
\geq\left\Vert \widetilde{w}(i)-\widetilde{w}(j)\right\Vert _{\infty}-\left\Vert \bar{w}^{1}-\widetilde{w}(i)\right\Vert _{\infty}-\left\Vert \bar{w}^{2}-\widetilde{w}(j)\right\Vert _{\infty}\\
\geq\left\Vert \widetilde{w}(i)-\widetilde{w}(j)\right\Vert _{\infty}-2\lambda.
\end{array}\label{eq:cv2_pr_2}
\end{equation}
Also, from the Lipschitz continuity property of $\mathfrak{f}$ and
from (\ref{eq:meas_2}), we have that
\begin{equation}
\begin{array}{c}
\mathfrak{f}(\bar{w}^{1})-\mathfrak{f}(\bar{w}^{2})\leq\mathfrak{f}(\widetilde{w}(i))-\mathfrak{f}(\widetilde{w}(j))+2\gamma_{\mathfrak{f}}\lambda\\
=\widetilde{z}(i)-e(i)-\widetilde{z}(j)+e(j)+2\gamma_{\mathfrak{f}}\lambda\\
\leq\widetilde{z}(i)-\widetilde{z}(j)-2\varepsilon+2\lambda+2\gamma_{\mathfrak{f}}\lambda
\end{array}\label{eq:cv2_pr_3}
\end{equation}
where the last inequality follows from (\ref{eq:cv2_pr_1}). Considering
that $\mathfrak{f}(\bar{w}^{1})>\mathfrak{f}(\bar{w}^{2})$, inequalities
(\ref{eq:cv2_pr_2}) and (\ref{eq:cv2_pr_3}) imply that
\[
\begin{array}{c}
\dfrac{\left|\mathfrak{f}(\bar{w}^{1})-\mathfrak{f}(\bar{w}^{2})\right|}{\left\Vert \bar{w}^{1}-\bar{w}^{2}\right\Vert _{\infty}}=\dfrac{\mathfrak{f}(\bar{w}^{1})-\mathfrak{f}(\bar{w}^{2})}{\left\Vert \bar{w}^{1}-\bar{w}^{2}\right\Vert _{\infty}}\leq\dfrac{\widetilde{z}(i)-\widetilde{z}(j)-2\varepsilon+2\lambda+2\gamma_{\mathfrak{f}}\lambda}{\left\Vert \widetilde{w}(i)-\widetilde{w}(j)\right\Vert _{\infty}-2\lambda}.\end{array}
\]
Since $\lambda$ is arbitrarily small, we have that, as $N\rightarrow\infty$,
\[
\frac{\left|\mathfrak{f}(\bar{w}^{1})-\mathfrak{f}(\bar{w}^{2})\right|}{\left\Vert \bar{w}^{1}-\bar{w}^{2}\right\Vert _{\infty}}\leq\frac{\widetilde{z}(i)-\widetilde{z}(j)-2\varepsilon}{\left\Vert \widetilde{w}(i)-\widetilde{w}(j)\right\Vert _{\infty}}.
\]
But $\frac{\left|\mathfrak{f}(\bar{w}^{1})-\mathfrak{f}(\bar{w}^{2})\right|}{\left\Vert \bar{w}^{1}-\bar{w}^{2}\right\Vert _{\infty}}=\gamma_{\mathfrak{f}}$
and $\frac{\widetilde{z}(i)-\widetilde{z}(j)-2\varepsilon}{\left\Vert \widetilde{w}(i)-\widetilde{w}(j)\right\Vert _{\infty}}\leq\hat{\gamma}$,
where it has been considered that, from Theorem \ref{thm:conv1},
$\lim\limits _{N\rightarrow\infty}\hat{\varepsilon}=\varepsilon$.
It follows that, as $N\rightarrow\infty$,
\begin{equation}
\gamma_{\mathfrak{f}}\leq\hat{\gamma}.\label{eq:cv2_pr_4}
\end{equation}
On the other hand, since $\left|e(k)\right|\leq\varepsilon,\:\forall k$,
\[
\begin{array}{c}
\tilde{\gamma}_{ij}=\dfrac{\widetilde{z}(i)-\widetilde{z}(j)-2\varepsilon}{\left\Vert \widetilde{w}(i)-\widetilde{w}(j)\right\Vert _{\infty}}\leq\dfrac{\mathfrak{f}(\widetilde{w}(i))-\mathfrak{f}(\widetilde{w}(j))+e(i)-e(j)-2\varepsilon}{\left\Vert \widetilde{w}(i)-\widetilde{w}(j)\right\Vert _{\infty}}\leq\dfrac{\mathfrak{f}(\widetilde{w}(i))-\mathfrak{f}(\widetilde{w}(j))}{\left\Vert \widetilde{w}(i)-\widetilde{w}(j)\right\Vert _{\infty}}\leq\gamma_{\mathfrak{f}},\:\forall i,j.\end{array}
\]
It follows that $\gamma_{\mathfrak{f}}\geq\hat{\gamma}$, which, together
with (\ref{eq:cv2_pr_4}), proves the claim.

\emph{Proof of Theorem \ref{lip_conv}}. Following the same lines
of the proof of Theorem 3 in \cite{NoFaMiAUT13}, it can be shown
that
\[
\limsup_{N\rightarrow\infty}\gamma_{\Delta}\leq\gamma_{\Delta}^{s}<\dfrac{1-\hat{\gamma}_{g,y}}{\hat{\gamma}_{f}}
\]
where $\hat{\gamma}_{f}$ and $\hat{\gamma}_{g,y}$ are estimates
of the Lipschitz constants $\gamma_{f}$ and $\gamma_{g,y}$ in (\ref{E:unif_cont_f})
and (\ref{E:unif_cont_y}), see step \ref{st_gg}) of Algorithm \ref{des_alg}.

The claim follows from Assumptions \ref{A:system_cont}-\ref{A:controller_cont},
\ref{A:noise}-\ref{dense_ass-2} and Theorem \ref{thm:conv1}.

\end{document}